\documentclass[12pt]{iopart}

\usepackage{color}
\usepackage[dvipsnames]{xcolor}
\usepackage{graphicx}
\usepackage{bm}
\usepackage{amsbsy}
\usepackage{hyperref}
\usepackage{enumerate}
\usepackage{upgreek}
\usepackage[subnum]{cases}
\usepackage{dsfont}

\newcommand{\ti}{ \tilde }
\newcommand{\ep}{ \epsilon }
\newcommand{\pa}{ \partial }
\newcommand{\hb}{ \hbar }
\newcommand{\si}{ \sigma }

\newcommand{\ga}{ \gamma }
\newcommand{\la}{ \langle }
\newcommand{\ra}{ \rangle }

\newcommand{\al}{ \alpha }

\newcommand{\Ga}{ \Gamma }

\newcommand{\cl}{ \mbox{cl} }
\newcommand{\pr}{\mbox{Pr}}
\newcommand{\fr}{\mbox{free}}
\newcommand{\re}{ \mbox{Re} }
\newcommand{\im}{ \mbox{Im} }

\newcommand{\erfc}{ \mbox{erfc} }

\newcommand{\lam}{ \lambda }
\newcommand{\Lam}{ \Lambda }
\newcommand{\thb}{ \tilde{\hbar} }
\newcommand{\tpsi}{ \tilde{\psi} }
\newcommand{\tphi}{ \tilde{\phi} }
\newcommand{\tj}{ \tilde{j} }
\newcommand{\trho}{ \tilde{\rho} }

\newcommand{\tPsi}{ \tilde{\Psi} }
\newcommand{\tPhi}{ \tilde{\Phi} }
\newcommand{\tJ}{ \tilde{J} }
\newcommand{\tDel}{ \tilde{\Delta} }

\newcommand{\ttb}{ \tilde{t}_b }
\newcommand{\tN}{ \tilde{N} }
\newcommand{\tSi}{ \tilde{\Sigma}_T }
\newcommand{\tD}{ \tilde{D}_1 }
\newcommand{\tDD}{ \tilde{D}_2(T) }
\newcommand{\tG}{ \tilde{G} }
\newcommand{\tg}{ \tilde{g} }

\newcommand{\tuo}{ \tilde{u}_0 }

\begin{document}

\title[Different routes to the classical limit of backflow]{Different routes to the classical limit of backflow}

\author{S. V. Mousavi}
\address{Department of Physics, University of Qom, Ghadir Blvd., Qom 371614-6611, Iran}
\ead{vmousavi@qom.ac.ir}
\author{S. Miret-Art\'es}
\address{Instituto de F\'isica Fundamental, Consejo Superior de Investigaciones Cient\'ificas, Serrano 123, 28006 Madrid, Spain}
\ead{s.miret@iff.csic.es}

\vspace{10pt}
\begin{indented}
\item[]October 2022
\end{indented}

\begin{abstract}

Decoherence is a well established process for the emergence of classical mechanics in open quantum systems. However,  it can have two different origins or mechanisms depending on the dynamics one is considering, speaking then  about intrinsic decoherence for isolated systems and environmental decoherence due to dissipation/fluctuations for open systems. This second mechanism can not be considered for backflow since no thermal fluctuation terms can be added in the formalism in order to keep an important requirement for the occurrence of this effect: only contributions of  positive momenta along time should be maintained. The purpose of this work is to analyze the backflow effect in the light of the underlying intrinsic decoherence and the dissipative dynamics. For this goal, we first deal with  the Milburn approach where a mean frequency of the unitary evolution steps undergone for the system is assumed. A comparative analysis is carried out in terms of the Lindblad master equation. Second, the so-called quantum-to-classical transition wave equation is analyzed  from a linear scaled Schr\"odinger equation which is derived and expressed in terms of a continuous parameter covering from the quantum to the classical regime as well as all in-between dynamical non-classical regimes. This theoretical analysis is inspired by the Wentzel-Kramers-Brillouin  approximation. And third, in order to complete our analysis, the transition wave equation formalism is also applied to dissipative backflow within the Caldirola-Kanai approach where the dissipative dynamics comes from an effective Hamiltonian. In all the cases treated here, backflow is gradually suppressed as the intrinsic decoherence process  is developing, paying a special attention to the classical limit. The route to classicality is not unique.

\end{abstract}

%
\vspace{2pc}

\section{Introduction}

Backflow is usually referred when free particles described by a one-dimensional wave
function consisting of a distribution of positive momenta along time, display an increasing probability of remaining in the negative region during certain periods of time (backflow intervals).
This counter-intuitive quantum phenomenon was initially noticed 
by Allock \cite{allcock} in his study of arrival times in quantum mechanics. A detailed study of the problem was only carried out 
25 years later by Bracken and Melloy \cite{BrMe-JPA-1994} where they introduced a new quantum number, today known 
as the Bracken-Melloy constant, being independent on the backflow interval, particle mass and Planck constant. This value was  computed 
numerically to be around 0.04 and is the least upper bound for backflow i.e., the greatest amount of probability for flowing from right to left through  
the origin. Furthermore, it appears to be the largest eigenvalue of a homogeneous Fredholm integral equation of second kind. 
Some rigorous results about the backflow operator have been derived elsewhere \cite{PeGrKrWa-JPA-2006}.

This effect has also been studied in the presence of a constant force  \cite{BM-APL-1998} and  extended to the 
realm of relativistic quantum mechanics \cite{MeBr-FP-1998, SuCh-MPLA-2018, AsLySt-PS-2019, BiBiAu-JPA-2022}. Its connection to 
arrival times \cite{YeHa-JPC-2013, GrKrPeRu-ACP-2006}, diffraction in time \cite{Gu-PRA-2019} and quantum reentry 
\cite{Gu-PRR-2020} have  been considered. In this regard, it has been demonstrated that construction of correlated quantum states 
for which the amount of backflow can exceed the Bracken-Melloy constant is possible \cite{Gu-PRR-2020}. 
More recently, a new formulation of the quantum backflow problem for arbitrary momentum distributions has been 
proposed and applied for gravitational and harmonic potentials \cite{Quantum1,Quantum2}. Moreover, backflow has been considered for the rotational 
motion \cite{Euro,Pra1,Pra2,Pra3} and has been addressed in connection with nonclassicality and the Wigner function \cite{IJQI}. 
Studies of quantum backflow have also been  extended to open quantum systems 
within the Caldirola-Kanai (CK) framework \cite{MoMi-EPJP-2020} as well as
for identical particles  \cite{MoMi-RP-2020}. A general formulation of quantum backflow for conservative systems of free non-relativistic 
many particles, both distinguishable and identical, has been carried out \cite{Ba-PRA-2020}.
While  experimental evidences for this effect have not  been reported yet, its optical counterpart has recently been  
observed \cite{ElZaBa-Opt-2020}.

Decoherence or the emergence of classical properties in a given physical system is usually considered in the context of open 
quantum dynamics where the environment is entangled with the system of interest. This kind of decoherence is usually called 
external or environmental decoherence  \cite{Sc-PR-2019,St-PTRSA-2012}. Just in the opposite side it is found in the literature what is
known as intrinsic decoherence which is intrinsic to Nature itself (it is ineluctable)  and where no environment is involved 
\cite{Mi-PRA-1991, St-PTRSA-2012} and, therefore, it can even exist for an isolated system. 
Over the years, several different paths have been taken to break up with the standard quantum mechanics; some of the well-known attempts include the postulation of extra nonlinear terms or stochastic terms in the Schr\"odinger equation; and gravitational effects invalidates this equation \cite{St-PTRSA-2012}. 
Several years ago, Milburn \cite{Mi-PRA-1991}  postulated  that, at sufficiently short time steps, 
the system does not evolve  continuously under unitary evolution but  by a stochastic sequence of identical unitary transformations.  
This means that  a minimum time step in the universe is introduced. The inverse of this time step is the mean frequency of the unitary 
steps. The  system is then considered isolated and described by a state whose evolution is governed by a generalized von Neumann equation having
this mean frequency as a parameter.

In this work, we also propose an additional intrinsic decoherence inspired by the Wentzel-Kramers-Brillouin (WKB) approximation.
In an attempt to describe both the quantum and classical regimes in a uniform language and  continuous way, a non-linear 
wave equation has been  proposed for such a transition  \cite{RiScVaBa-PRA-2014}. It has been 
then proved that this non-linear equation is equivalent to a linear one which is just a Schr\"odinger equation but with a scaled 
Planck constant and scaled wave function \cite{MoMi-AP-2018,MoMi-JPC-2018}. 
This transition wave equation contains a  parameter ranging from one  (quantum regime) to zero (classical regime) covering 
all the regimes in-between (non-classical dynamical regimes) which is equivalent to consider the limit $\hbar \rightarrow 0$. 
The procedure of using a continuous parameter monitoring the different dynamical regimes in the theory could be seen quite similar 
to the WKB approximation (based on a series expansion in powers of Planck constant), widely used for conservative systems in the
energy domain. However, some important differences should be clearly stressed.
First, whereas the classical Hamilton-Jacobi equation for the  action is obtained at zero order in the WKB approximation,  
the so-called classical time-dependent (non-linear) wave equation \cite{Schi-PR-1962} is reached by construction. 
Second, the hierarchy of the differential equations for the action at different orders of the expansion in $\hbar$ is substituted by  a
transition wave differential equation which can be  solved in the linear and non-linear domains. Third, 
the transition from quantum to classical trajectories is carried out in a continuous and gradual way, covering all the intermediate 
non-classical dynamical regimes. Fourth, the scaling procedure extended and applied to open quantum systems is very easy to implement.
And fifth, the gradual (intrinsic) decoherence process due to the scaled Planck constant allows us to analyze the continuity of the 
intermediate dynamical regimes to finally reach the classical regime.
At this level, it is interesting to point out that this process studied within this theoretical scheme displays  decoherence due to  
the gradual transition from the quantum to the classical regime. 
This kind of studies has been extended to dissipative and stochastic dynamics in the framework of the CK and the so-called 
Schr\"odinger-Langevin or Kostin equations, respectively \cite{MoMi-AP-2018, MoMi-JPC-2018,MoMi-FOP-2022}.

Finally, an interesting aspect deserving special attention is the classical limit of the backflow effect; in particular, 
because of the independence of the Bracken-Melloy constant from the Planck constant, this limit cannot be taken naively. 
An attempt has already been carried out by Yearsley et al. \cite{YeHaHaWh-PRA-2012, YeHa-JPC-2013}  using 
quasi-projectors in the definition of the flux operator. Recently, Bracken \cite{Br-PS-2021} 
has proposed another way for treating the classical limit of backflow by generalizing the eigenvalue problem where the allowed range 
of momentum values is expanded beyond those previously considered. 
In this work, our purpose is also to consider this goal by analyzing it in terms of the intrinsic decoherence and dissipative dynamics.
Such a study in the direction of the classical limit of backflow via the scaled Schr\"odinger equation confirms 
that backflow is a non-classical effect in the sense that it is not observed in systems of classical particles. However, it should be 
noted that this effect is a wave phenomenon which can be seen for classical waves too \cite{BiBiAu-JPA-2022}. This is due to the fact that
the wave-particle duality is absent in classical mechanics. The situation is similar to quantum tunneling but in classical optics; 
frustrated total reflection \cite{Krane-book-2012}.   
Two particular problems will be analyzed, the Bracken and Melloy classical example  both for non-dissipative and dissipative 
dynamics is studied and the free propagation of a superposition of two Gaussian wave packets. 
In all of these cases, backflow is gradually suppressed as the intrinsic decoherence process is 
developing and the classical limit is reached.

\section{Backflow within the von Neumann formalism}

In the most general formulation of quantum mechanics, a physical system is described by a density operator
$\hat{\rho}$ instead of a state vector $ | \psi \rangle$. In this context, the von Neumann equation
\begin{eqnarray} \label{eq: Neumann}
	i \hb \frac{\pa \hat{\rho}}{\pa t} &=& [\hat{H}, \hat{\rho}]
\end{eqnarray}
has to be applied for the time evolution of the system, $ \hat{H} $ being the Hamiltonian of the system. For a single particle and in one dimension, 
this Hamiltonian is expressed as 
\begin{eqnarray}
	\hat{H} &=& \frac{ \hat{p}^2 }{2m} + V( \hat{x} )   ,
\end{eqnarray}
where the first term is the kinetic energy operator and the second term, the external interaction potential. We are interested in backflow for free particles. For this goal, the Hamiltonian is reduced to the first term describing the kinetic energy of free particles.

In the momentum representation, the von Neumann equation (\ref{eq: Neumann}) for free particles is written as
\begin{eqnarray} \label{eq: Neumann_momentum}
	i \hb \frac{\pa}{\pa t} \rho(p, p', t) &=& \frac{p^2-p'^2}{2m} \rho(p, p', t)    ,
\end{eqnarray}
its  solution being
\begin{eqnarray} \label{eq: Neumann_sol}
	\rho(p, p', t) &=& \exp \left[ - \frac{i}{\hb} \frac{p^2-p'^2}{2m} t \right] \rho(p, p', 0)    .
\end{eqnarray}
On the contrary, in the position representation,  Eq. (\ref{eq: Neumann}) is expressed as
\begin{eqnarray} \label{eq: Neumann_coor}
	i \hb \frac{\pa}{\pa t} \varrho(x, x', t) &=& -\frac{ \hb^2 }{2m} \left( \frac{\pa^2}{\pa x^2} - \frac{\pa^2}{\pa x'^2} \right) \varrho(x, x', t)   ,
\end{eqnarray}
which, in terms of the center of mass and relative coordinates, $ R = (x+x')/2 $ and $ r=x-x' $, it can be rewritten as
\begin{eqnarray} \label{eq: Neumann_coor2}
	\frac{\pa}{\pa t} \varrho(R, r, t) + \frac{\pa}{\pa R} j(R, r, t)  &=& 0   ,
\end{eqnarray}
where we have introduced the probability current density matrix 
\begin{eqnarray} \label{eq: curmat}
	j(R, r, t)  &=& \frac{\hb}{i m} \frac{\pa}{\pa r} \varrho(R, r, t)  .
\end{eqnarray}
Diagonal elements  are obtained from $r=0$ and $R=x$,  $ \varrho(x, t) = \varrho(R=x, r=0, t)$ and $ j(x, t) = j(R=x, r=0, t) $ being then the probability density and probability current density fulfilling the continuity equation 
\begin{eqnarray} \label{eq: Neumann_continuity}
	\frac{\pa}{\pa t} \varrho(x, t) + \frac{\pa}{\pa x} j(x, t)  &=& 0.
\end{eqnarray}

From the continuity equation (\ref{eq: Neumann_continuity}), 
one has, after integration over $x$ between $- \infty$ and $0$, that
\begin{eqnarray} \label{eq: Pt}
	\frac{d}{dt} \pr(t) &=& - j(0, t)    ,
\end{eqnarray}
where $ \pr(t) $ gives the probability of finding the particle in the negative half-space, $ x < 0 $; we have assumed that $ j(-\infty, t) = 0 $. 
Except if $j(0, t)$ is negative, $ \pr(t)$ is a decreasing function of time.
If $j(0, 0)$ is negative then, by continuity in time, $j(0, t) $ will be negative over some time interval, say  $ [0, t_b) $ \cite{BrMe-JPA-1994}. 
Then, from Eq. (\ref{eq: Pt}), the increasing probability  or transmitted right-to-left probability in this time interval is given by 
\begin{eqnarray} \label{eq: Delta}
	{\Delta}_{t_b} &=&  \pr(t_b) - \pr(0) = - \int_0^{t_b} dt ~ j(0, t)
	= \int_0^{t_b} dt ~ j_{_-}(0, t)   ,
\end{eqnarray}
where in the last equality we have used the negativity of $j$ in the interval $ [0, t_b) $ and introduced
\begin{eqnarray} \label{eq: jneg}
	j_{_-}(0, t) &=& \frac{ | j(0, t) | - j(0, t) }{2}   .
\end{eqnarray}
%
%
If the contribution of negative momenta to the probability density at all times is zero, then one speaks of backflow and the interval $ [0, t_b) $ is called the backflow time interval.

The density matrix in the position representation written in terms of its elements in the momentum space are computed from
\begin{eqnarray}
	\varrho(x, x', t) &=& \la x | \hat{\rho}(t) | x' \ra = \int_{-\infty}^{\infty}dp  \int_{-\infty}^{\infty}dp'
	\la x | p \ra \la p | \hat{\rho}(t) | p' \ra \la p' | x' \ra \nonumber \\
	&=& \frac{1}{2\pi \hb} \int_{-\infty}^{\infty}dp  \int_{-\infty}^{\infty}dp'
	e^{i(px-p'x')/\hb} ~ e^{ -i (p^2-p'^2)t/2m\hb } \rho(p, p', 0)   ,  \label{eq: rho_pos}
\end{eqnarray}
where in the second line we have used Eq. (\ref{eq: Neumann_sol}).  
From Eqs. (\ref{eq: rho_pos}) and (\ref{eq: curmat}), we have then that
\begin{eqnarray}
	j(0, t) &=& \frac{1}{4m\pi \hb} \int_{-\infty}^{\infty}dp  \int_{-\infty}^{\infty}dp' (p+p') e^{ -i (p^2-p'^2)t/2m\hb } \rho(p, p', 0)  .
\end{eqnarray}
%
For a mixture of states i.e.,
\begin{eqnarray}
	\rho(p, p', 0) &=& \sum_i w_i \phi_i(p) \phi_i^*(p'), \qquad \sum_i w_i = 1   ,
\end{eqnarray}
$j$ is also a mixture with the same weights $w_i$,
\begin{eqnarray}
	j(0, t) &=& \sum_i w_i j_i(0, t),
\end{eqnarray}
with
\begin{eqnarray} \label{eq: ji}
	j_i(0, t) &=& \frac{1}{4m\pi \hb} \int_{-\infty}^{\infty}dp  \int_{-\infty}^{\infty}dp' (p+p') e^{ -i (p^2-p'^2)t/2m\hb } \phi_i(p) \phi_i^*(p')     .
\end{eqnarray}
For quantum backflow, only contributions coming from positive momenta
will be considered.

As an illustration, let us consider the mixture of two states $ \phi_1(p) $ and $ \phi_2(p) $ with weights $w$ and $1-w$, respectively, where
\begin{eqnarray}
	\phi_1(p) &=& \frac{18}{ \sqrt{35 ~ \hb k} } \frac{p}{ \hb k } \left( e^{-p / \hb k} - \frac{1}{6} e^{-p / 2\hb k} \right) \Theta(p) \label{eq: BM-state}  \\
	\phi_2(p) &=&\sqrt{ \frac{2}{\hb k} } e^{-p / \hb k} \Theta(p)   ,
\end{eqnarray}
$k$ being a positive wave number and $\Theta(p)$ is the Heaviside function. 
Eq. (\ref{eq: BM-state}) is the state used by Bracken and Melloy in their first study of backflow \cite{BrMe-JPA-1994}.
Then, from Eq. (\ref{eq: ji}) we have that 
\begin{eqnarray}
	j_1(0, 0) &=& - \frac{36}{35 \pi} \frac{\hb k^2}{m} , \\
	j_2(0, 0) &=& \frac{1}{\pi} \frac{\hb k^2}{m} , \\
	j(0, 0) &=& w j_1(0, 0) + (1-w)j_2(0, 0) = \frac{1}{\pi} \frac{\hb k^2}{m} \left( 1 - w \frac{71}{35} \right)  .
\end{eqnarray}
These results show that the state $ \phi_1(p) \phi_1^*(p') $ displays backflow while the state $ \phi_2(p) \phi_2^*(p') $ does not. Thus,
the mixed state $ \rho(p, p', 0) = w \phi_1(p) \phi_1^*(p') + (1-w) \phi_2(p) \phi_2^*(p') $ displays quantum backflow only when
$ 35 / 71 < w \leq 1 $.

\section{Backflow in the framework of the Milburn approach}

Several years ago, Milburn \cite{Mi-PRA-1991} introduced what is known as  {\it intrinsic decoherence} in quantum mechanics by  
postulating  that, at sufficiently short time steps, the system does not evolve  continuously under unitary evolution but  by a 
stochastic sequence of identical unitary transformations.  This means that  a minimum time step in the universe should be  introduced. 
The inverse of  this time step is the mean frequency of the unitary steps, $\lam$. 
In this way, a generalized evolution equation was proposed whose expansion to first order in $\lam^{-1}$ reads as
\begin{eqnarray} \label{eq: Milburn}
	\frac{\pa \hat{\rho}}{\pa t} &=& - \frac{i}{\hb} [\hat{H}, \hat{\rho}] - \frac{1}{2\hb^2 \lam} [\hat{H}, [\hat{H}, \hat{\rho}] ]   .
\end{eqnarray}
In the limit $ \lam \rightarrow \infty $, this equation reduces to the von Neumann equation (\ref{eq: Neumann}). Furthermore, the 
Milburn equation (\ref{eq: Milburn}) preserves the trace of the density operator.


In the momentum representation, the Milburn equation (\ref{eq: Milburn}) for free particles reads as
\begin{eqnarray} \label{eq: mil-mom}
	\frac{\pa}{\pa t} \rho(p, p', t) &=&  \left[ - \frac{i}{\hb}\frac{p^2-p'^2}{2m} - \frac{1}{2\hb^2 \lam} \left( \frac{p^2-p'^2}{2m} \right)^2 \right] \rho(p, p', t)   ,
\end{eqnarray}
its solution being
\begin{eqnarray} \label{eq: mil_sol}
	\rho(p, p', t) &=& \exp \left[ - \frac{i}{\hb} \frac{p^2-p'^2}{2m} t - \frac{1}{2\hb^2 \lam} \left( \frac{p^2-p'^2}{2m} \right)^2  t \right] \rho(p, p', 0)   .
\end{eqnarray}
As this solution shows, diagonal elements are constant in time while non-diagonal elements decay exponentially  showing the effect of the 
so-called {\it intrinsic decoherence}. For environmental decoherence, within the Caldeira-Leggett formalism, the non-diagonal terms decay with time
at a rate governed by the damping constant and temperature \cite{KhMoMi-En-2021}.
Furthermore, if the initial state contains only positive momenta, the corresponding time evolving state too. 

Eq. (\ref{eq: Milburn}) for free particles in the position representation takes now the form
\begin{eqnarray} 
\frac{\pa}{\pa t} \varrho(x, x', t) &=& 
\left[ - \frac{i}{\hb} - \frac{1}{2\hb^2 \lam} \frac{ - \hb^2 }{2m} \left( \frac{\pa^2}{\pa x^2} - \frac{\pa^2}{\pa x'^2} \right)
\right]
\frac{ - \hb^2 }{2m} \left( \frac{\pa^2}{\pa x^2} - \frac{\pa^2}{\pa x'^2} \right) \varrho(x, x', t)   .
\label{eq: mil-pos} \nonumber \\
\end{eqnarray}
In the $(R, r)$ coordinates, this equation is equivalent to Eq. (\ref{eq: Neumann_coor2}) but with the probability current density matrix given by
\begin{eqnarray} \label{eq: mil-curmat}
	j(R, r, t)  &=& \left( - \frac{ i \hb}{m} + \frac{ \hb^2 }{2 \lam m^2} \frac{\pa^2}{\pa r \pa R}\right) \frac{\pa}{\pa r} \varrho(R, r, t)      .
\end{eqnarray}
The density matrix in the position representation is then given by the inverse Fourier transform of its representation in the momentum space according to
\begin{eqnarray} \label{eq: rho_pos_mil}
	\varrho(R, r, t) &=& \frac{1}{2\pi \hb} \int_{-\infty}^{\infty}dp  \int_{-\infty}^{\infty}dp'
	e^{i(p-p')R/\hb} e^{i(p+p')r/2\hb} \rho(p, p', t)    ,
\end{eqnarray}
and using Eq. (\ref{eq: mil-curmat}) the probability current density is given by
\begin{eqnarray} \label{eq: mil-jd}
	j(x, t)  &=& \frac{1}{4\pi m \hb} \int_{-\infty}^{\infty}dp  \int_{-\infty}^{\infty}dp' e^{i (p-p')x/\hb }
	(p+p') \left( 1 - i \frac{p^2-p'^2}{4\hb m \lam} \right) \rho(p, p', t) , \nonumber \\
\end{eqnarray}
which is a real function due to the fact that $ \rho(p, p', t) = \rho^*(p', p, t) $. In order to simplify the notation,
one can use some characteristic length $ \mu $ and time $ \nu = \mu^2 m / \hb $ and  then represent dimensionless variables by capital letters according to
\begin{eqnarray}
	X &=& \frac{x}{\mu}    , \label{eq: dimless_cor}\\
	T &=& \frac{ t }{\nu}   ,  \label{eq: dimless_time} \\
	P &=& \mu \frac{p}{ \hb }  , \label{eq: dimless_mom}
\end{eqnarray}
and for  functions,
\begin{eqnarray}	
	\Psi(X, T) &=& \sqrt{\mu} ~ \psi(x, t)   ,\\
	\Phi(P, T) &=& \sqrt{ \frac{\hb}{\mu} } ~ \phi(p, t)   , \label{eq: dimless_phi} \\ 
	J(X, T) &=& \nu ~ j(x, t)   . \label{eq: dimless_j}
\end{eqnarray}
Thus, one has that
\begin{eqnarray} \label{eq: mil-jd-dim}
	J(X, T)  &=& \frac{1}{4\pi} \int_{-\infty}^{\infty}dP  \int_{-\infty}^{\infty}dP' e^{i (P-P')X }
	(P+P') \left( 1 - i \frac{P^2-P'^2}{4\Lam} \right) \bar{\rho}(P, P', T)  , \nonumber \\
\end{eqnarray}
where $ \Lam = \nu \lam $ and $ \bar{\rho}(P, P', T) = (\hb/\mu)~\rho(p, p', t) $ and 
\begin{eqnarray} 
	J(0, T)  &=& \frac{1}{4\pi} \int dP  \int dP'
	(P+P') ~ \re \{\bar{\rho}(P, P', T) \} \nonumber \\
	&+& 
	\frac{1}{16\pi \Lam} \int dP  \int dP'
	(P+P')^2(P-P') ~ \im \{\bar{\rho}(P, P', T) \}   . \label{eq: mil-j00-dim}
\end{eqnarray}
Note that the standard quantum mechanics expression is reached when $ \Lam \rightarrow \infty $. Thus, the second term shows 
the contribution of the intrinsic decoherence in the backflow dynamics. 

We now consider two specific examples. As stated previously, for backflow the contribution of negative 
momenta to the state must be zero and $ J(0, 0) $ has to be negative.
%
%
\begin{figure}  
	\includegraphics[width=14cm,angle=-0]{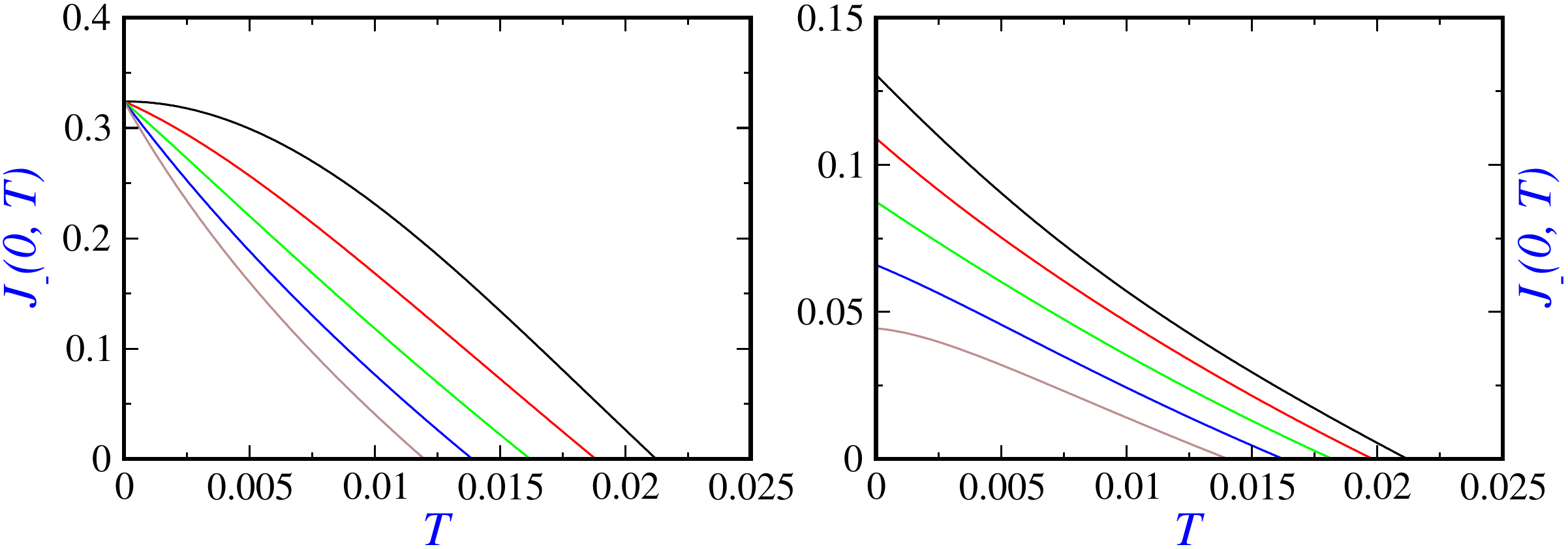}
	\caption{
		Negative part of the probability current density at the origin for different values of $\Lam^{-1}$: $ \Lam^{-1} = 0 $ (black), 
		$ \Lam^{-1} = 0.005 $ (red), $ \Lam^{-1} = 0.01 $ (green), 
		$ \Lam^{-1} = 0.015 $ (blue)  and $ \Lam^{-1} = 0.02 $ (brown) for the initial state given by Eq. (\ref{eq: mil-example1})  (left panel) and Eq. (\ref{eq: mil-example2}) (right panel).	}
	\label{fig: JnegMilburn}
\end{figure}
%
%
In figure \ref{fig: JnegMilburn},  the negative part of the probability current density at the origin, $ J_-(0, T) $, 
for different values of $ \Lam $ for a pure state is plotted
\begin{eqnarray} 
	\bar{\rho}(P, P', 0)  &=& \Phi(P) \Phi^*(P') 
\end{eqnarray}
with the so-called Bracken-Melloy state Eq. (\ref{eq: BM-state}) (left panel)
\begin{eqnarray} 
	\Phi(P) &=&  \frac{ 18 }{\sqrt{35 } } P 
	\left( e^{- P } - \frac{1}{6} e^{- P / 2 } \right) \Theta(P) \label{eq: mil-example1}
\end{eqnarray}
and a different state (right panel) given by
\begin{eqnarray} 
	\Phi(P) &=& \sqrt{ \frac{ 1823508 }{ 253055 } } P 
	\left( e^{- (1+i/2) P } - \frac{1}{6} e^{-(1+i/2) P / 2 } \right) \Theta(P)   . \label{eq: mil-example2}
\end{eqnarray}
Note that dimensionless space and time coordinates are used according to  
\begin{numcases}~
	\mu = \frac{1}{k} \label{eq: lenght-dim-BM} \\
	\nu = \frac{\mu^2 m}{\hb} = \frac{m}{\hb k^2}   . \label{eq: time-dim-BM} 
\end{numcases}
In the former case, the contribution of the second term in Eq. (\ref{eq: mil-j00-dim}) is zero initially whereas 
this is not true in the second case. Because of this, in the right panel, the curves corresponding to different values of $ \Lam $ start initially 
at different values. 
As this figure also shows, backflow is suppressed gradually as $ \Lam $ decreases revealing the role of the intrinsic decoherence. 
Note that both the backflow interval and the amount of backflow are reduced due to this type of decoherence.

\begin{figure}  
\centering
\includegraphics[width=12cm,angle=-0]{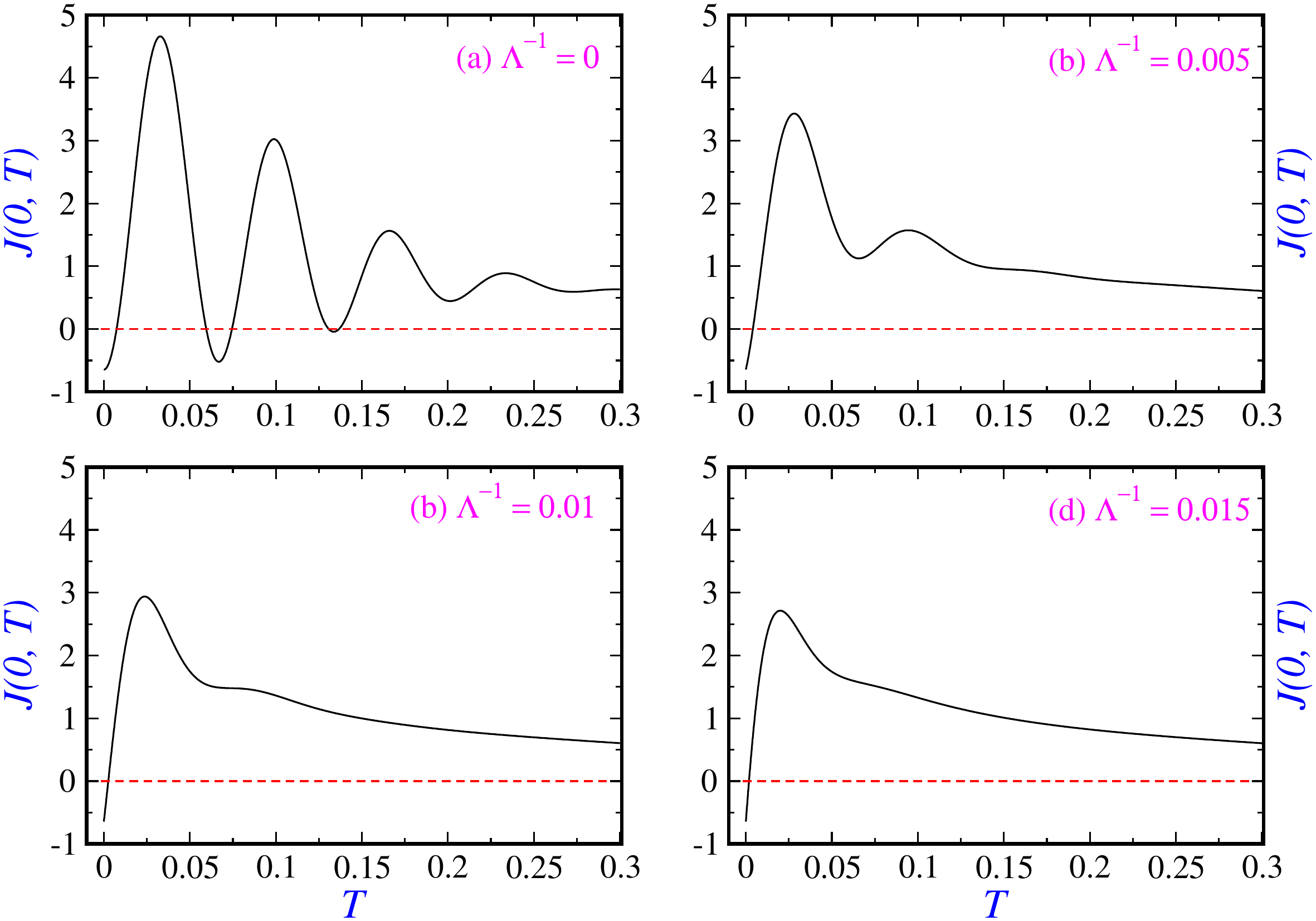}
\caption{
The probability current density at the origin versus time for different values of $\Lam^{-1}$ for an initial state given by the superposition of two 
Gaussian wave packets, Eq. (\ref{eq: mil-example3}) (see text for the parameters used).
}
\label{fig: curt_G}
\end{figure}

\begin{figure}  
\centering
\includegraphics[width=12cm,angle=-0]{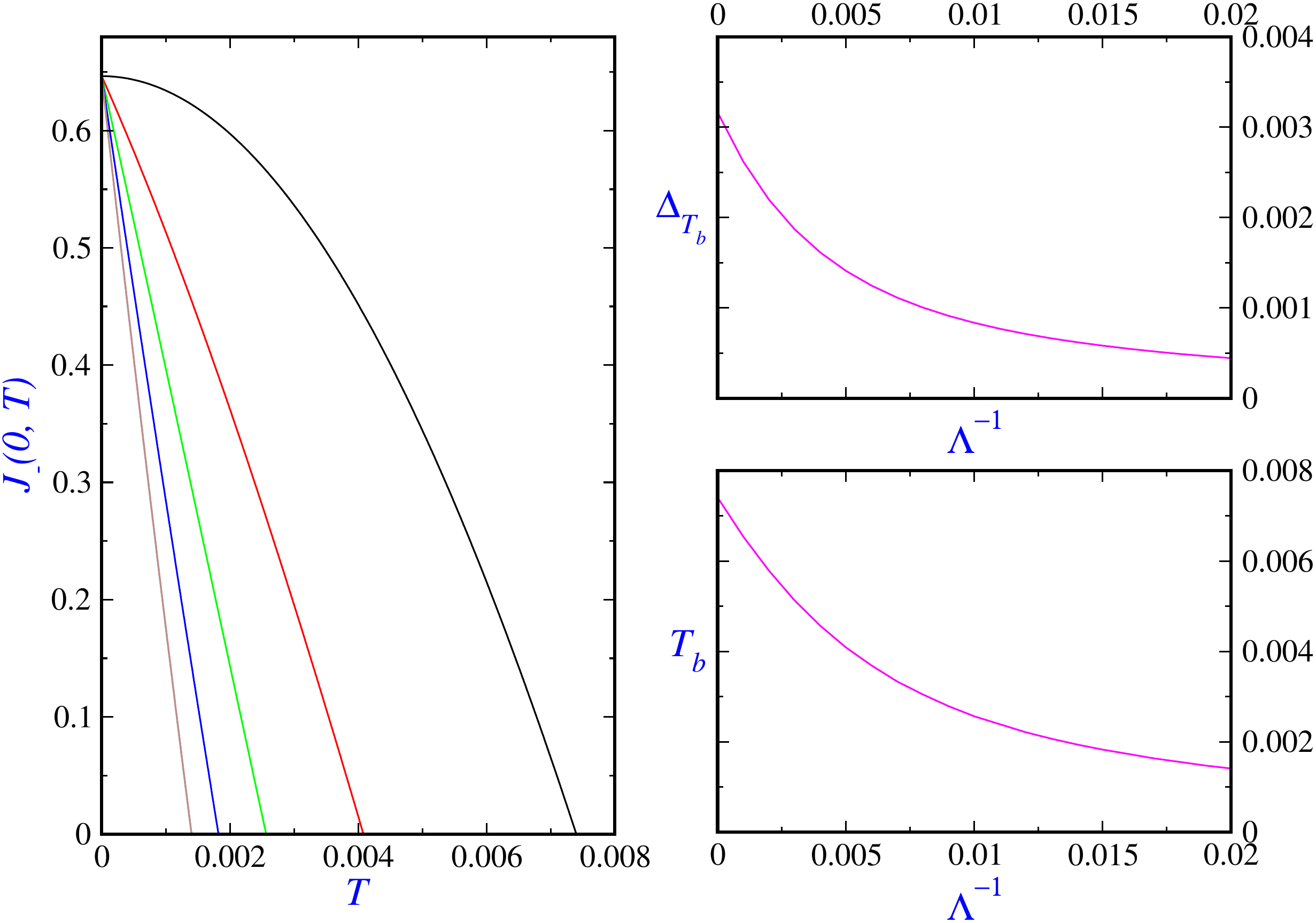}
\caption{
Negative part of the probability current density at the origin (left panel) for different values of $\Lam^{-1}$: $ \Lam^{-1} = 0 $ (black), 
$ \Lam^{-1} = 0.005 $ (red), $ \Lam^{-1} = 0.01 $ (green), $ \Lam^{-1} = 0.015 $ (blue)  and $ \Lam^{-1} = 0.02 $ (brown).  
In the right two panels, the amount of backflow (top) and the first backflow duration (bottom) for the superposition of two Gaussian 
wave packets (\ref{eq: mil-example3}) as a function of $\Lam^{-1}$.
}
\label{fig: jneg_G}
\end{figure}

In figure \ref{fig: curt_G}, the probability current density at the origin is plotted versus time for different values of the parameter $ \Lambda^{-1} $ for the superposition of two Gaussian wave packets in the momentum representation
\begin{eqnarray} 
\Phi(P) &=& N \left( \frac{2}{\pi} \right)^{1/4} \left\{ e^{-(P-P_{0a})^2} + \al e^{i\theta} e^{-(P-P_{0b})^2} \right\} , \label{eq: mil-example3} 
\end{eqnarray}
where the normalization factor is given by
\begin{eqnarray}
N &=& \left( 1 + \al^2 + 2 \al \cos \theta ~ e^{-(P_{0a}-P_{0b})^2/2} \right)^{-1/2}   .
\end{eqnarray}
Here space and time have been expressed in terms of
\begin{numcases}~
\mu = \frac{\hb}{2\si_p} \label{eq: lenght-dim-Gauss} \\
\nu = \frac{\mu^2 m}{\hb} = \frac{m \hb}{4 \si_p^2}, \label{eq: time-dim-Gauss} 
\end{numcases}
$ \si_p $ being width of the component wavepackets. 
With values $P_{0a} = 14$, $P_{0b}=3$, $\al = 1.9$ and $\theta=\pi$ the contribution of negative momenta to the wave packet is negligible around $ 10^{-10} $. 
As clearly seen, for the standard quantum mechanics where $ \Lambda^{-1} = 0 $ (panel a), there are three intervals of backflow while 
for the remaining three other values (panels b, c and d) only a single backflow interval is seen at very short times. Furthermore, the 
intrinsic decoherence makes this time interval shorter. This fact is better observed  in the left panel of figure \ref{fig: jneg_G} where 
the corresponding negative part is plotted
for different values of $\Lambda^{-1}$ and the right bottom panel where the duration of backflow is depicted in terms of $\Lambda^{-1}$. 
In the right panels of the same figure, the amount of backflow (top) quantified by means of 
the integral of negative part of  the probability current density over the first backflow interval $[0, T_b)$ 
\begin{eqnarray} \label{eq: beta}
\Delta_{T_b} &=& \int_0^{T_b} dt ~ J_-(0, t)
\end{eqnarray}
and the first backflow duration (bottom) are plotted as a function of $\Lambda^{-1}$.
One sees again that the intrinsic decoherence reduces the amount of backflow. 
%
%
%
%
%

\subsection{The Milburn and Lindblad equations}

A Lindbladian master equation has the form \cite{Sch-book-2007}
\begin{eqnarray} \label{eq: ld1}
	\frac{\pa}{\pa t} \hat{\rho}(t) &=& - \frac{i}{\hb} [\hat{H}', \hat{\rho}(t)] 
	- \frac{1}{2} \sum_{\al} \kappa_{\al} [\hat{L}_{\al}, [\hat{L}_{\al}, \hat{\rho}(t)]]   ,
\end{eqnarray}
$ \hat{H}' $ being the the renormalized (Lamb-shifted) Hamiltonian of the system, $ \hat{L}_{\al} $ the so-called Lindblad operators and  a family of parameters $ \kappa_{\al} \geq 0 $.
For free particles, the subject of our study, this master equation with just one Lindblad operator 
recasts
\begin{eqnarray} \label{eq: ld2}
	\frac{\pa}{\pa t} \hat{\rho}(t) &=& - \frac{i}{\hb} [\frac{ \hat{p}^2 }{2m}, \hat{\rho}(t)] 
	- \frac{1}{2} \kappa [\hat{L}, [\hat{L}, \hat{\rho}(t)]] .
\end{eqnarray}
%
%
%
Comparison of the Milburn equation with the power-expanded in inverse of the time step, $\lam^{-1}$,  
\begin{eqnarray} \label{eq: Milburn1}
	\frac{\pa}{\pa t} \hat{\rho}(t) &=& - \frac{i}{\hb} [\frac{ \hat{p}^2 }{2m}, \hat{\rho}(t)] - \frac{1}{2\hb^2 \lam} [\frac{ \hat{p}^2 }{2m}, [\frac{ \hat{p}^2 }{2m}, \hat{\rho}(t)] ] 
\end{eqnarray}
with the master equation (\ref{eq: ld2}) reveals that this equation, in its expanded form, is just a Lindbladian master equation with only a Lindblad operator proportional to $ \hat{p}^2 $. Because of this similarity, one motivates to consider backflow when the Lindblad operator is a different function of {\it only} the momentum.
To this end, one must first make sure that the contribution of negative momenta to the state remains zero along time.
Taking $ \hat{L} = f(\hat{p}) $, $f$ being an arbitrary well-defined function, from Eq. (\ref{eq: ld2}) in the momentum representation we have that 
\begin{eqnarray} \label{eq: ld3}
	\frac{\pa}{\pa t} \rho(p, p',t) &=& \left\{ - \frac{i}{\hb} \frac{ p^2-p'^2 }{2m}
	- \frac{1}{2} \kappa (f(p)-f(p'))^2 \right\}\rho(p, p',t)
\end{eqnarray}
whose solution reads as
\begin{eqnarray} \label{eq: solld3}
	\rho(p, p', t) &=& \exp \left[ \left( - \frac{i}{\hb} \frac{p^2-p'^2}{2m} - \frac{1}{2} \kappa (f(p)-f(p'))^2 \right) t \right] \rho(p, p', 0)  .
\end{eqnarray}
This equation confirms that the probability density $ \rho(p, p, t) $ is independent on time. Thus, if the contribution of negative momenta to the distribution is initially zero it will be so forever.  
By considering the special simple form $ \hat{L} = \hat{p} $, Eq. (\ref{eq: ld2}) is expressed as
\begin{eqnarray} \label{eq: ld_pos}
	\frac{\pa}{\pa t} \varrho(x, x', t) &=& 
	\left[ \frac{i \hb }{2m} \left( \frac{\pa^2}{\pa x^2} - \frac{\pa^2}{\pa x'^2} \right)
	+ \frac{1}{2} \kappa \hb^2 \left( \frac{\pa}{\pa x} + \frac{\pa}{\pa x'} \right)^2
	\right]
	\varrho(x, x', t) 
\end{eqnarray}
in the coordinate representation. Such a choice of the Lindblad operator has already been used to study the effect of dephasing in the context of neutron Compton scattering \cite{ChDrTi-JPCS-2010}.
In terms of the coordinates $ (R, r) $, Eq. (\ref{eq: ld_pos}) takes the form (\ref{eq: Neumann_coor2}) with the probability current density matrix 
\begin{eqnarray} \label{eq: curmat}
	j(R, r, t)  &=& \left( \frac{\hb}{i m} \frac{\pa}{\pa r} - \frac{1}{2} \kappa \hb^2 \frac{\pa}{\pa R} \right) \varrho(R, r, t)  .
\end{eqnarray}
Expressing $ \varrho(x, x', t) $ in terms of $ \rho(p, p', t) $ and using Eq. (\ref{eq: curmat}), then one obtains
\begin{eqnarray} \label{eq: PCD00}
	j(0, 0)  &=& \frac{1}{4m\pi \hb} \int_{-\infty}^{\infty}dp  \int_{-\infty}^{\infty}dp' 
	( p + p' - i m \hb \kappa  ) \rho(p, p', 0)
\end{eqnarray}
for the probability current density at the origin.

We have avoided the more standard decoherence mechanism in which the Lindblad operator is proportional to the position $ \hat{x} $ on the grounds that this is incompatible with keeping positive momentum. However, this deserves an in-depth investigation. Taking $ \hat{L} = \hat{x} $, Eq. (\ref{eq: ld2}) recasts
\begin{eqnarray} \label{eq: ld_pos-x}
	\frac{\pa}{\pa t} \varrho(x, x', t) &=& 
	\left[ \frac{i \hb }{2m} \left( \frac{\pa^2}{\pa x^2} - \frac{\pa^2}{\pa x'^2} \right)
	- \frac{1}{2} \kappa (x-x')^2 \right]
	\varrho(x, x', t) 
\end{eqnarray}
in the coordinate representation which is the equation of motion for environmental scattering \cite{Sch-book-2007}. On the other hand, the Caldeira-Leggett (CL) master equation in the high temperature limit reads as
\begin{eqnarray} \label{eq: CL eq}
	\frac{\pa}{\pa t}\varrho(x, x', t) &=& \left[\frac{i \hb}{2m} \left( \frac{\pa^2}{\pa x^2} - \frac{\pa^2}{\pa x'^2} \right) - \ga (x-x') \left( \frac{\pa}{\pa x} - \frac{\pa}{\pa x'} \right)
	- \frac{D}{\hb^2} (x-x')^2 \right] \varrho(x, x', t) \nonumber \\
\end{eqnarray}
where $\ga$ is the relaxation rate or damping constant and $ D = 2 m \ga k_B T_e $ plays the role of the diffusion coefficient 
($k_B$ is the Boltzmann constant and $T_e$ is the temperature of the environment). 
Comparison with Eq. (\ref{eq: ld_pos-x}) shows that the CL equation (\ref{eq: CL eq}) is a Lindbladian one only in the negligible dissipation limit where the second term can be ignored. 
%
Let us consider now the evolution of a single Gaussian under Eq. (\ref{eq: ld_pos-x}) where the contribution of negative momenta is addressed by computing the probability of obtaining negative momenta in a measurement.
The evolution of the Gaussian wavepacket 
\begin{eqnarray} \label{eq: 1gauss}
	\psi_0(x) &=& \frac{1}{ (2\pi \si_0^2)^{1/4} } \exp \left[ - \frac{x^2}{4\si_0^2} + i \frac{p_0}{\hb} x  \right]
\end{eqnarray}
under the Lindbladian master equation (\ref{eq: ld_pos-x}) yields \cite{MoMi-EPJP-2022}
\begin{eqnarray} \label{eq: sol ld_pos-x} 
	\varrho(R, r, t) &=& \frac{1}{ \sqrt{2\pi a_2(t)} } \exp\left[ a_0(r, t) - \frac{ ( R - a_1(r, t) )^2 }{ 4 a_2(t)} \right] ,
\end{eqnarray}
where
\begin{eqnarray}
	a_0(r, t)& =& - \left( \frac{ 1}{ 8 \si_0^2 } + \frac{\kappa}{2}  t \right) r^2
	+ i  \frac{p_0}{\hb} r \equiv  - a_{02}(t) ~ r^2 + i ~ a_{01} ~ r ,
	\label{eq: a0}
	\\
	a_1(r, t) &=& \frac{p_0}{m} t + i \hb \left( \frac{ 1 }{ 4 m \si_0^2 }t + \frac{\kappa}{2m} t^2 \right) r \equiv a_{10}(t) 
	+ i~ a_{11}(t) ~ r ,
	\label{eq: a1} \\ 
	a_2(t) &=& \frac{\si_0^2}{2} \left( 1 + \frac{ \hb^2 }{ 4 m^2 \si_0^4 } t^2
	+ \frac{ \hb^2 \kappa }{3 m^2 \si_0^2} t^3 \right)   .
	\label{eq: a2} 
\end{eqnarray}
The argument of the exponential function in Eq. (\ref{eq: sol ld_pos-x}) can be written as 
\begin{equation} \label{eq: arg_exp}
	- \frac{ (R - a_{10}(t))^2 }{4a_2(t)} + i \frac{ R a_{11}(t) - a_{10}(t) a_{11}(t) + 2 a_{01} a_2(t)}{2a_2(t)} r - \left( a_{02}(t) - \frac{a_{11}(t)^2}{4 a_2(t)}  \right) r^2
\end{equation}
in powers of $r$ from which one obtains widths in the diagonal, $x=x'$, and off-diagonal, $x=-x'$, directions respectively as
\begin{eqnarray}
	\si_t &=& \sqrt{2 a_2(t)} = \si_0 \sqrt{1 + \frac{ \hb^2 }{ 4 m^2 \si_0^4 } t^2
		+ \frac{ \hb^2 \kappa }{3 m^2 \si_0^2} t^3}  ,
	\label{eq: st} \\
	\ell(t) &=& \left\{ 8 \left( a_{02}(t) - \frac{a_{11}(t)^2}{4 a_2(t)}  \right) \right \}^{-1/2}  ,
\end{eqnarray}
where $ \si_t $ is width of the probability density $ \varrho(R=x, r=0, t) $ and the so-called coherence length $\ell(t)$ measures the characteristic distance over which the system can exhibit spatial interference effects \cite{Sch-book-2007}. The short time behaviour of the coherence length is given by
\begin{eqnarray} 
	\ell(t) &\approx& \si_0 \left[ 1 - 2 \si_0^2 \kappa ~ t + \left( \frac{\hb^2}{8m^2\si_0^4} + 6 \si_0^4 \kappa^2  \right)t^2  \right] . \label{eq: ell-short} 
\end{eqnarray}
From this, one can extract the decoherence time-scale to be
\begin{eqnarray} \label{eq: deco-time}
	\tau_{\mbox{decoh}} \sim \frac{1}{2 \si_0^2 \kappa}   .
\end{eqnarray}
Schlosshauer \cite{Sch-book-2007} takes this time-scale as the localization time-scale while it should be instead $ \sqrt{m/(\hb \kappa)} $ (see for instance Ref. \cite{HaZo-PRD-1997}.) 
There is another way to extract the decoherence time. The time evolution of the density matrix $ \varrho(x, x', t) $ under only the non-unitary term $ - \frac{1}{2} \kappa (x-x')^2 $ in Eq. (\ref{eq: ld_pos-x}) yields
\begin{eqnarray} \label{eq: nu-evolv}
	\varrho(x, x', t) &=& \varrho(x, x', 0) \exp \left[ - \frac{1}{2} \kappa (x-x')^2 ~ t \right]   .
\end{eqnarray}
This shows that the suppression of off-diagonal terms (coherences) increases exponentially with time and with the
squared separation $ (\Delta x)^2 = (x - x')^2 $. From Eq. (\ref{eq: nu-evolv}), one defines a decoherence time-scale as
\begin{eqnarray} \label{eq: deco-time}
	\tau_{\Delta x} = \frac{2}{\kappa (\Delta x)^2 }  .
\end{eqnarray}
For an initial superposition of localized states a distance $ b $ apart, the decoherence time-scale has been obtained to be $ 4 / (\kappa b^2) $.

\begin{figure} 
	\centering
	\includegraphics[width=12cm,angle=-0]{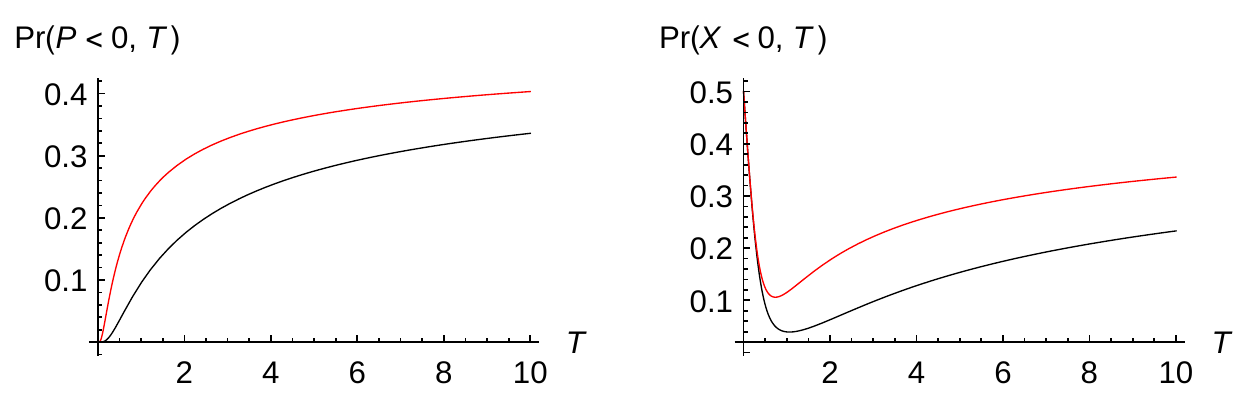}
	\caption{
		$ \pr(P<0, T) =  \erfc [ \sqrt{2} P_0 / \sqrt{1 + 4 \bar{\kappa} T} ]/2 $ (right panel) and $ \pr(X<0, T) =  \erfc [ P_0 T / (\sqrt{2} \sqrt{1 + T^2/4 + \bar{\kappa} T^3 / 3}) ]/2 $ (left panel) for $P_0=3$ for $ \bar{\kappa} = 5 $ (black) and $ \bar{\kappa} = 15 $ (red). For units of length, time and $ \kappa $ we have used respectively $\si_0$, $ \si_0^2 m / \hb $ and $ \hb / ( m \si_0^4 ) $. Note that for $P_0=3$ the  probability for obtaining negative values in a measurement of momentum given by Eq. (\ref{eq: pr_pneg_mom}) is {\it initially} of the order of $ 10^{-10} $.
	}
	\label{fig: pxnegppneg} 
\end{figure}

Now the evolution in the momentum space is carried out by means of the Fourier transform of  Eq. (\ref{eq: 1gauss}) which gives
\begin{eqnarray} \label{eq: 1gauss_mom}
	\phi_0(x) &=& \left(\frac{2\si_0^2}{ \pi \hb^2 }\right)^{1/4} \exp \left[ -\frac{\si_0^2}{\hb^2} (p-p_0)^2  \right]
\end{eqnarray}
and leads to \cite{KhMoMi-En-2021}
\begin{eqnarray} \label{eq: pd_g_mom}
	\rho(p, p, t) &=& \frac{1}{\sqrt{2\pi}w_t} \exp \left[ - \frac{(p-p_0)^2}{2 w_t^2} \right], \qquad 
	w_t = \frac{\hb}{2\si_0} \sqrt{1+4 \kappa \si_0^2 t}  ,
\end{eqnarray}
for the probability density from which one obtains
\begin{eqnarray} \label{eq: pr_pneg_mom}
	\pr(p<0, t) &=& \frac{1}{2} \erfc \left[ \frac{p_0}{ \sqrt{2} w_t } \right]   ,
\end{eqnarray}
for the probability of obtaining a negative value for momentum. 
The width $w_t$ increases with time. Thus, the argument of the complementary error function in Eq. (\ref{eq: pr_pneg_mom}) is a decreasing function of time revealing that the probability $ \pr(p<0, t) $ is an increasing function of time.
In this way, we have that $ \tau_a \sim 1 / (4 \kappa \si_0^2) $ is the  time where negative momenta start appearing, taking $ p_0 \si_0$ at the same order of $\hb$.
These negative momenta affect the probability of finding the particle in the negative half-space
\begin{eqnarray} \label{eq: pr_xneg}
	\pr(x<0, t) &=& \int_{-\infty}^0 dx' \varrho(x, x, t) = \frac{1}{2} \erfc \left[ \frac{(p_0/m)t}{ \sqrt{2} \si_t } \right]  .
\end{eqnarray}
The argument of this complementary error function becomes a decreasing function of time for times greater than $ \tau_s = (6 m^2 \si_0^2 / (\hb^2 \kappa) )^{1/3} $. This is the time where the effect of negative momenta is reflected in the position. In figure \ref{fig: pxnegppneg}, both the probability of obtaining a negative value in a measurement of momentum (left panel) and  the probability of finding the particle in the negative half-space (right panel) are plotted. As this figure shows, $ \tau_s $ is at least one order of magnitude greater than $ \tau_a $.

\section{ Backflow in the framework of the quantum to classical transition wave equation }

\subsection{Preliminaries}

The so-called classical non-linear Schr\"odinger equation is written as \cite{Ro-AJP-1964}
\begin{eqnarray} \label{eq: Class-Sch}
	i \hbar \frac{\partial \psi_{\cl}(x, t)}{\partial t} &=& \left[ -\frac{\hbar^2}{2m} \frac{\partial^2}{\partial x^2} + V(x) + \frac{\hbar^2}{2m} \frac{1}{\mid \psi_{\cl}(x, t) \mid} \frac{\partial^2 \mid \psi_{\cl}(x, t) \mid}{\partial x^2}  \right] \psi_{\cl}(x, t) ,  \nonumber \\
\end{eqnarray}
which can be obtained from the  quantum to classical  transition wave equation
\begin{eqnarray} \label{eq: tran}
	i \hb \frac{\partial}{\partial t}\psi_{\ep}(x, t) &=& \left[ -\frac{\hb^2}{2m}  \frac{\pa^2}{\pa x^2} + V(x) + (1-\ep)\frac{\hb^2} {2m \mid \psi_{\ep}(x, t) \mid} \frac{\partial^2  \mid \psi_{\ep}(x, t) \mid}{\partial x^2} 
	\right] \psi_{\ep}(x, t) . \nonumber \\
\end{eqnarray}
This non-linear transition wave equation was proposed for a continuous transition from the quantum to the classical 
regime \cite{RiScVaBa-PRA-2014} and has been proven to be equivalent to the so-called  linear scaled wave 
equation \cite{RiScVaBa-PRA-2014} 
\begin{eqnarray} \label{eq: Scaled Sch}
	i \ti{\hb} \frac{\pa}{\pa t}\ti{\psi}(x, t) &=& \left[ -\frac{\ti{\hb}^2}{2m} \frac{\pa^2}{\pa x^2} + V(x) 
	\right] \ti{\psi}(x, t)    ,
\end{eqnarray}
where
\begin{eqnarray} \label{eq: scaled Planck}
	\ti{\hb} &=& \hbar ~ \sqrt{\ep} 
\end{eqnarray}
is the scaled Planck constant with $\ep$ being a continuous parameter going  from unity (quantum regime) to zero (classical regime). The scaled wave function $	\ti{\psi}(x, t) $ is written in
terms of the transition wave function $ \psi_{\ep}(x, t) $ as
\begin{eqnarray} \label{eq: psitilde-psiep}
	\ti{\psi}(x, t) &=& \psi_{\ep}(x, t) \exp \left[ \frac{i}{\hb} \left( \frac{1}{\sqrt{\ep}} - 1  \right) S_{\ep}(x, t) \right] ,
\end{eqnarray}
with $ S_{\ep}(x, t) $ being the phase of the transition wave function. 
For an ensemble of classical particles, positions are distributed according to Born rule $ \mid \psi_{\cl} \mid^2 $ 
while momenta are uniquely determined via $ p = \partial S_{cl} / \partial x $; $ S_{cl} $ being the  phase of the classical wave function.
In this context, the $\ep$ parameter is going to play a similar role as $\lambda$ in the Milburn approach. The intrinsic decoherence process 
is ruled now by $\ep$ when going from one to zero.

From the linear scaled Schr\"odinger equation (\ref{eq: Scaled Sch}), one obtains the continuity equation
\begin{eqnarray} \label{eq: con_eq}
	\frac{\pa \trho }{\pa t} + \frac{\pa \tj }{\pa x} &=& 0    ,
\end{eqnarray}
where the scaled probability density and probability current density are 
\begin{numcases}~ 
	\trho(x, t) = | \tpsi(x, t) |^2   \label{eq: pd} , \\
	\tj(x, t) = \frac{\thb}{m} \im \left\{ \tpsi^*(x, t) \frac{\pa}{\pa x} \tpsi(x, t) \right \}    , \label{eq: pcd}
\end{numcases}
respectively. Thus, from Eq. (\ref{eq: con_eq}),  one again finds, after integration 
over the negative part of the space and assuming $ \tj(-\infty, t) = 0 $, that
\begin{eqnarray} \label{eq: dPt_dt}
	\frac{d}{dt} \ti{P}(t) &=& - \tj(0, t)    ,
\end{eqnarray}
where $ \ti{P}(t) $ is the scaled probability of finding the particle in the negative half-space $ x < 0 $. 
If $ \tj(0, 0)$ is negative, then, by continuity in time, $ \tj(0, t) $ will be negative over some time interval, say $ [0, \ttb) $ \cite{BrMe-JPA-1994}. 
Similarly to previous sections, the  right-to-left transmitted probability in this time interval is given by 
\begin{eqnarray} \label{eq: Delta}
	\ti{\Delta}_{\ttb} &=&  \ti{P}(\ttb) - \ti{P}(0) = - \int_0^{\ttb} dt ~ \tj(0, t)
	= \int_0^{\ttb} dt ~  \tj_{_-}(0, t)   ,
\end{eqnarray}
by introducing again  
\begin{eqnarray} \label{eq: jneg}
	\tj_{_-}(0, t) &=& \frac{ | \tj(0, t) | - \tj(0, t) }{2}   
\end{eqnarray}
%
%
in terms of the corresponding scaled function.
If the contribution of positive momenta to the wavefunction at all times is zero, then one speaks of backflow and the interval $ [0, \ttb) $ is  the backflow interval.

As before, we are going to work on dimensionless variables. By using the same definitions as given by Eqs. (\ref{eq: dimless_cor}), 
(\ref{eq: dimless_time}), and (\ref{eq: dimless_mom}), and adding a few more for the scaled functions
%
\begin{eqnarray}
	\tPsi(X, T) &=& \sqrt{\mu} ~ \tpsi(x, t)    , \\
	\tPhi(P, T) &=& \sqrt{ \frac{\hb}{\mu} } ~ \tphi(p, t)    , \label{eq: dimless_phi} \\ 
	\tJ(X, T) &=& \nu ~ \tj(x, t)    , \label{eq: dimless_j}\\
	\tG(X, X'; T) &=& \mu ~ \tg(x, x'; t)   ,
\end{eqnarray}
%
the  coordinate-space scaled wave function 
\begin{eqnarray}
	\tpsi(x, t) &=& \int_{-\infty}^{\infty} dx' ~ \tg(x, x'; t) ~ \tpsi(x',0)   
\end{eqnarray}
is expressed now as
\begin{eqnarray} \label{eq: dimless_wf}
	\tPsi(X, T) &=& \int_{-\infty}^{\infty} dX' ~ \tG(X, X'; T) ~ \tPsi(X',0) 
\end{eqnarray}
for the dimensionless wave function. 
For {\it free} particles, the corresponding wave function can also be obtained from the momentum space wave function as follows
\begin{eqnarray} \label{eq: wf_invfour}
	\tpsi(x, t) &=& \frac{1}{\sqrt{2\pi \thb}} \int dp ~ e^{i( px - p^2 t/2m)/\thb} \tphi(p)   ,
\end{eqnarray}
or
\begin{eqnarray} \label{eq: Freewf}
	\tPsi(X, T) &=& \frac{1}{\ep^{1/4}} \frac{1}{\sqrt{2\pi}} \int dP ~ e^{i( PX - P^2 T/2)/ \sqrt{\ep}} ~ \tPhi(P)    .
\end{eqnarray}

\subsection{Backflow as an eigenvalue problem}

Bracken and Melloy \cite{BrMe-JPA-1994} showed that the highest probability which can flow back 
from positive to negatives values of the coordinate is around 0.04.
%
The maximum amount of backflow probability  occurring in general over any finite time interval is 
independent on the time interval, mass of the particle and Planck constant. A new dimensionless quantum number was then proposed. Recently, Bracken \cite{Br-PS-2021}, in an effort to partially solving the 
classical limit of backflow, has generalized the study to wave functions whose momentum contributions is within an arbitrary
interval instead of the original one $ [0, \infty) $.  
Here, in this subsection, we are going to carry out the same study but in the context of the scaled Schr\"odinger equation by considering 
the momentum contribution in the interval $ [p_0, \infty) $; $p_0$ being an arbitrary non-negative number.
Using Eqs. (\ref{eq: wf_invfour}) and (\ref{eq: Delta}), one obtains
\begin{eqnarray} \label{eq: Del-ker}
	\tDel_{\ttb} &=&  -\frac{1}{\pi} \int_{p_0}^{\infty} \int_{p_0}^{\infty} dp~dq~ \tphi^*(p) \frac{ \sin[(p^2 - q^2) \ttb / 4m\thb] }{p-q}~ e^{ i ( p^2-q^2 ) \ttb /4m\thb } \tphi(q)   . \nonumber \\
\end{eqnarray}
%
%
%
Now to find the optimum value of the right-to-left transported probability  $ \tDel_{\ttb} $, one should optimize the integral of 
Eq. (\ref{eq: Del-ker}) subject to the normalization condition
\begin{eqnarray}
	\int_{p_0}^{\infty} dp |\tphi(p)|^2 &=& 1 ,
\end{eqnarray}
by using  the method of Lagrange multipliers. After some straightforward algebra, one sees that the only Lagrange multiplier equals $ \tDel_{\ttb} $ and appears as the eigenvalue of the integral equation 
\begin{eqnarray} \label{eq: eigenvalue1}
	- \frac{1}{\pi} \int_{p_0}^{\infty} \frac{ \sin[(p^2-q^2)\ttb/4m\thb] }{p-q} \tphi(q) &=& \tDel(p_0) ~ \tphi(p)    ,
\end{eqnarray}
where, for simplicity, we have dropped the sub-index $ \ttb $ and for clarity we have introduced the right-to-left transported probability as a function of $ p_0 $. By setting
\begin{numcases}~
	p = 2\sqrt {\frac{ m \thb }{ \ttb }} ~ u   \label{eq: u}  \\
	q = 2\sqrt {\frac{ m \thb }{ \ttb }} ~ v \label{eq: v}  \\
	\phi( p ) = e^{iu^2} \varphi(u) \label{eq: phiu} \\
	\phi( q ) = e^{iv^2}\varphi(v) \label{eq: phiv}
\end{numcases}~
the eigenvalue equation (\ref{eq: eigenvalue1}) can be written as
\begin{eqnarray} \label{eq: eigenvalue2}
	- \frac{1}{\pi} \int_{\tuo}^{\infty} dv ~ \frac{ \sin(u^2-v^2) }{ u-v } \varphi(v)
	&=&  \tDel(\tuo) \varphi(u)    .
\end{eqnarray}
where $ \tuo = p_0 \sqrt{ \ttb / 4 m \hb \sqrt{\ep}  }  $.

%
\begin{figure}  
	\centering
	\includegraphics[width=12cm,angle=-0]{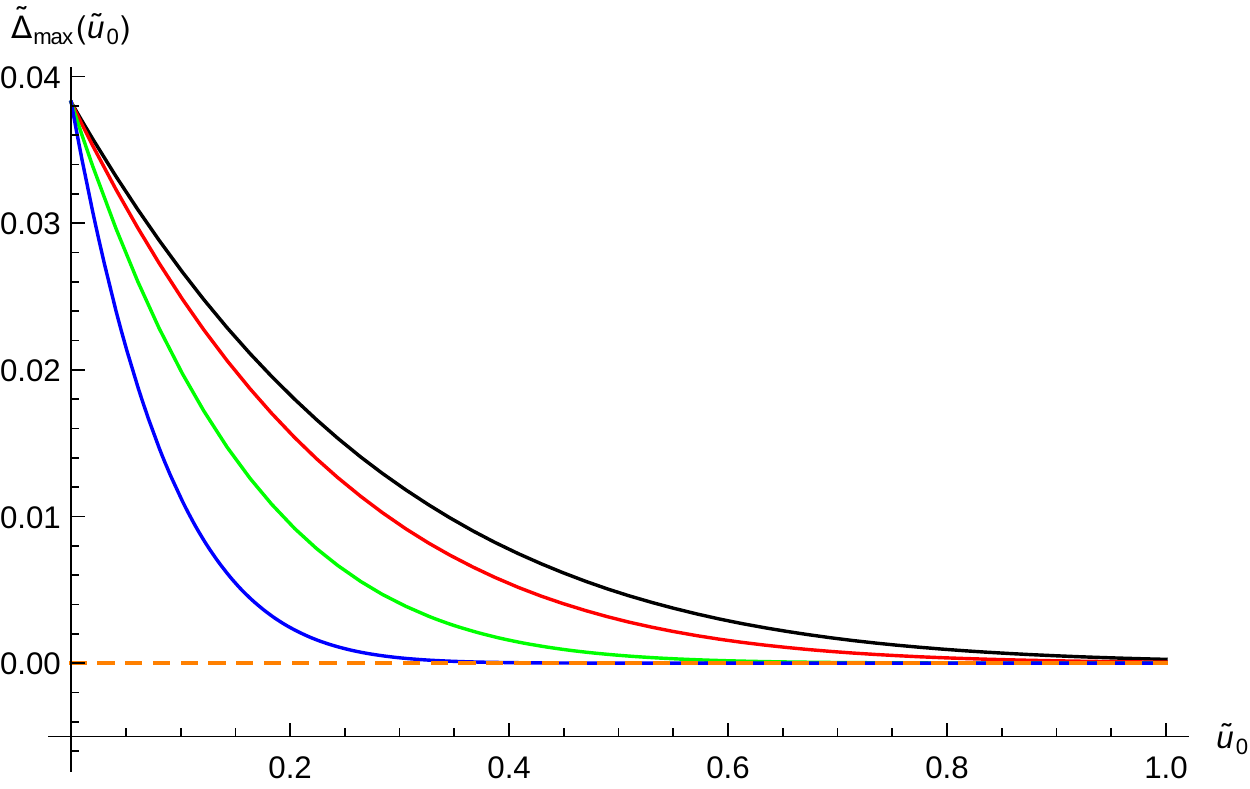}
	\caption{Maximum eigenvalue $ \tDel_{\max} $ versus $ \ti{u}_0 $ for different dynamical regimes: quantum (black), non-classical with 
		$ \ep = 0.5 $ (red), $ \ep = 0.1 $ (green) and $ \ep = 0.01 $ (blue) and classical (dashed orange). 	}
	\label{fig: pDelta} 
\end{figure}
%

In figure \ref{fig: pDelta} we have plotted the maximum backflow probability $ \tDel_{\max}(\tuo) $ versus $ \tuo $ for different 
dynamical regimes keeping the same backflow interval $ \ttb $. As clearly seen, for all the dynamical regimes,
$ \tDel_{\max}(\tuo) $ decreases with $ \tuo $. Furthermore, in the classical limit $ \ep \rightarrow 0 $, this probability approaches zero 
irrespective of $ \tuo $.

%

\subsection{Bracken and Melloy classical example}
Let us now consider a free particle which is described initially by the normalized wave function \cite{BrMe-JPA-1994}
\begin{eqnarray} \label{eq: phi_BM}
	\tphi(p) &=& \frac{18}{ \sqrt{35 ~ \thb k} } \frac{p}{ \thb k } \left( e^{-p / \thb k} - \frac{1}{6} e^{-p / 2\thb k} 
	\right) \Theta(p)   ,
\end{eqnarray}
in the momentum space where $ k $ is a positive wave number.
Expressing space and time coordinates from Eqs. (\ref{eq: lenght-dim-BM}) and (\ref{eq: time-dim-BM})
%
%
and using Eqs. (\ref{eq: dimless_phi}), (\ref{eq: dimless_j}) and (\ref{eq: Freewf}), one has that
\begin{numcases}~
	\tPhi(P) = \frac{1}{ \ep^{3/4} } \frac{18}{ \sqrt{35} } P \left( e^{-P / \sqrt{\ep} } - \frac{1}{6} e^{-P / (2 \sqrt{\ep}) } \right) \Theta(P) \label{eq: Phi_BM} \\
	\tPsi(X, T) = \frac{1}{ \ep^{1/4} } \frac{1}{\sqrt{2\pi} } \int dP ~ e^{i( PX - P^2 T/2)/ \sqrt{\ep} } ~ \tPhi(P)\\
	\tJ(X, T) = \sqrt{\ep} ~ \im \left\{ \tPsi^*(X, T) \frac{\pa}{\pa X} \tPsi(X, T) \right \} \label{eq: PCD-BM}
\end{numcases}
and
\begin{eqnarray} \label{eq: J00_BM}
	\tJ(0, 0) &=& - \sqrt{\ep} \frac{36}{35 \pi}
\end{eqnarray}
which is explicitly negative for all non-classical regimes and revealing the occurrence of backflow for all regimes except the classical one. 
This effect will not be certainly observed for the classical regime $ \ep = 0 $ since $ J_{cl}(0, 0) = 0 $. 
%
\begin{figure}  
	\centering
	\includegraphics[width=12cm,angle=-0]{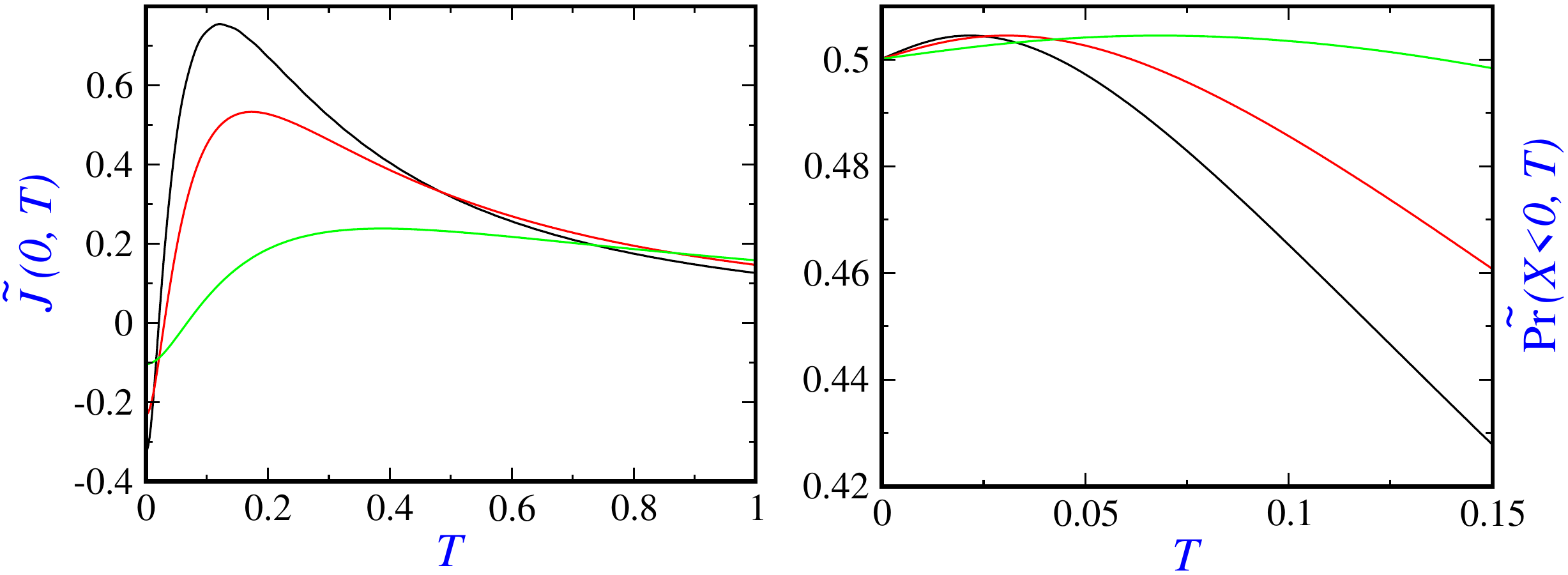}
	\caption{
		The probability current density given by Eq. (\ref{eq: PCD-BM}) (left panel) and the probability of remaining in the negative 
		half-space (right panel) for  different dynamical regimes: quantum (black) and non-classical (red and green with $\ep=0.5$ 
		and $\ep=0.1$, respectively).
	}
	\label{fig: plotjProb} 
\end{figure}
%
%
As figure \ref{fig: plotjProb} displays, the backflow interval increases while the negativity of the probability current density decreases 
in the quantum-to-classical transition. As should be expected, computations show that Eq. (\ref{eq: Delta}) is almost the same 
for all non-classical regimes; $ \tDel \simeq 0.004251 $. 

Several years ago, Halliwell {\em et al.} \cite{Halliwell2013} considered quantum backflow states from eigenstates of the so-called regularized current operator. They used a general function of momenta subject to special conditions, leading to a surprisingly large backflow of about 41 $\%$ of the lower bound on flux derived by Bracken and Melloy.

\subsection{A superposition of two Gaussian wave packets}

As an another illustration of backflow, we have solved the scaled Schr\"odinger equation (\ref{eq: Scaled Sch}) for free particles 
by considering an initial wave function as a superposition of two co-centred Gaussian wave packets with the same width but different kick momenta
\begin{eqnarray} \label{eq: wf_sup_0_Gauss}
	\tpsi_0(x) &=& \tN( \tpsi_a(x, 0) + \al e^{i\theta} \tpsi_b(x, 0) ) \nonumber \\
	& = & 
	\tN \frac{1}{( 2\pi \si_0^2 )^{1/4}} (e^{i p_{0a} x/\thb} + \al e^{i\theta} e^{i p_{0b} x/\thb} )~ e^{- x^2/4 \si_0^2}  ,
\end{eqnarray}
with
\begin{eqnarray} \label{eq: N_cons}
	\tN &=& \left( 1 + \al^2 + 2 \al e^{- \si_0^2 (p_{0a}-p_{0b})^2 / 2 \thb^2} \cos \theta \right)^{-1/2}    ,
\end{eqnarray}
and $\al$ and $\theta$ being two arbitrary real numbers.
Here space and time coordinates are expressed in terms of 
\begin{numcases}~
	\mu = \si_0   , \\
	\nu = \frac{\mu^2 m}{\hb} = \frac{m \si_0^2}{\hb}  .
\end{numcases}
The solution of the scaled Schr\"odinger equation (\ref{eq: Scaled Sch}) yields \cite{MoMi-EPJP-2020}
\begin{eqnarray} \label{eq: psi2_t}
	| \tPsi(X, T) |^2 & = & \tN^2 \frac{1}{\sqrt{2\pi} \tSi}
	\bigg\{
	\exp \left[ - \frac{ ( X - P_{0a} T )^2  }{ 2 \tSi^2 } \right] 
	+ \alpha \exp \left[ \tD - \frac{ ( X - \tDD )^2  }{ 2 \tSi^2 } \right]  \nonumber \\
	&+ & \alpha \exp \left[ \tD^* - \frac{ ( X - \tilde{D}_2^*(t) )^2  }{ 2 \tSi^2 } \right] 
+ \alpha^2 \exp \left[ - \frac{ ( X - P_{0b} T )^2  }{ 2 \tSi^2 } \right] 
	\bigg \}
\end{eqnarray}
for the probability density, where 
\begin{eqnarray} \label{eq: sigmat}
	\tSi &=& \sqrt{ 1 + \ep ~ T^2 / 4 }
\end{eqnarray}
is the width of each wave packet and
%
%
%
\begin{numcases}~
	\tD = - i \theta - \frac{ (P_{0a} - P_{0b})^2 }{2 \ep} \\
	\tDD = \frac{(P_{0a} + P_{0b}) T}{2}   + i \frac{ P_{0a} - P_{0b} }{ \sqrt{\ep} } \\
	\tN = \left( 1 + \al^2 + 2 \al e^{ - (P_{0a}-P_{0b})^2 / 2 \ep} \cos \theta \right)^{-1/2}    .
\end{numcases}
The momentum space wave function can be obtained as follows
\begin{eqnarray}
	\tphi(p, t) &=& \la p | \tpsi(t) \ra = \la p | e^{- i \hat{H} t / \thb} |\tpsi_0 \ra = 
	e^{- i p^2 t / 2 m \thb}  \tphi_0(p)   ,
\end{eqnarray}
where $  \tphi_0(p) $ is the Fourier transform of  $ \tpsi_0(x) $. With dimensionless variables, one has that
\begin{eqnarray} \label{eq: wf0_momentum}
	\tPhi_0(P) &=& \tN \frac{1}{ (2\pi)^{1/4} } \frac{ \sqrt{2} }{ \ep^{1/4} } \left\{
	\exp \left[ - \frac{ ( P-P_{0a} )^2}{\ep} \right]
	+ \al e^{i\theta} \exp \left[ - \frac{ ( P - P_{0b} )^2}{\ep} \right]
	\right \}.   \nonumber \\
\end{eqnarray}
Thus, the probability for obtaining a negative value in a measurement of momentum is time-independent and given by
\begin{eqnarray} \label{eq: prob_p<0}
\tilde{\pr}(p<0) &=& \int_{-\infty}^0 dp ~| \tphi(p, t) |^2 = \int_{-\infty}^0 dP ~| \tPhi(P, T) |^2 
\nonumber \\
&=& 
\frac{\tN^2}{2}  \bigg\{ 
\erfc \left[  \frac{ \sqrt{2} P_{0a} }{ \sqrt{\ep} } \right] 
+ \alpha^2 \erfc \left[ \frac{ \sqrt{2} P_{0b} }{ \sqrt{\ep} } \right]
\nonumber \\
&+& 2 \al ~e^{ - (P_{0a}-P_{0b})^2 / 2\ep } \cos \theta  ~ \erfc \left[ \frac{(P_{0a}+P_{0b})}{ \sqrt{2 \ep} } \right] \bigg\}     .
\end{eqnarray}
For backflow, one should make sure that the contribution of negative momenta to the wavefunction is zero. To this end, the parameters of 
the problem should be chosen in a way that $ \tilde{\pr}(p<0) $ is negligible. Two points are in order:
(i) the width of each Gaussian wave packets in the momentum space is $ \ti{\si}_p = \thb /2 \si_0 $ which becomes zero in the 
classical regime $ \ep = 0 $. 
(i)) Eq. (\ref{eq: prob_p<0}) shows that $ \tilde{\pr}(p<0) $ is zero in the classical regime $ \ep = 0 $ apart from the values of the involved parameters.

From the probability density Eq. (\ref{eq: psi2_t}), one obtains
\begin{eqnarray} \label{eq: prob_x<0}
\ti{ \pr }(X<0, T) &=& \frac{1}{2} \tN^2 \bigg\{ 
\erfc \left[ P_{0a} T / \sqrt{2} \tSi  \right]
+ \al^2 \erfc \left[ P_{0b} T / \sqrt{2} \tSi \right]
\nonumber \\
&+& 
2 \al 
e^{ - (P_{0a}-P_{0b})^2 / 2\ep } 
\bigg( \cos\theta~ \re \left\{ \erfc \left[ \tDD / ( \sqrt{2}  ~ \tSi ) \right]
\right\} 
\nonumber \\
&+&
\sin\theta~ \im \left\{ \erfc \left[ \tDD / ( \sqrt{2}  ~ \tSi ) \right]
\right\} \bigg)
\bigg \}
\end{eqnarray}
for the probability of remaining in the negative half-space $ x < 0 $. 
%
	%
	%
	%
Eq. (\ref{eq: prob_x<0}) shows that, in the classical regime, the contribution of interference terms becomes zero and the first two terms 
are decreasing functions of time. Thus, backflow is not observed in the classical realm while it is so in the remaining  non-classical regimes.
	%
	
\begin{figure}  
		\centering
		\includegraphics[width=12cm,angle=-0]{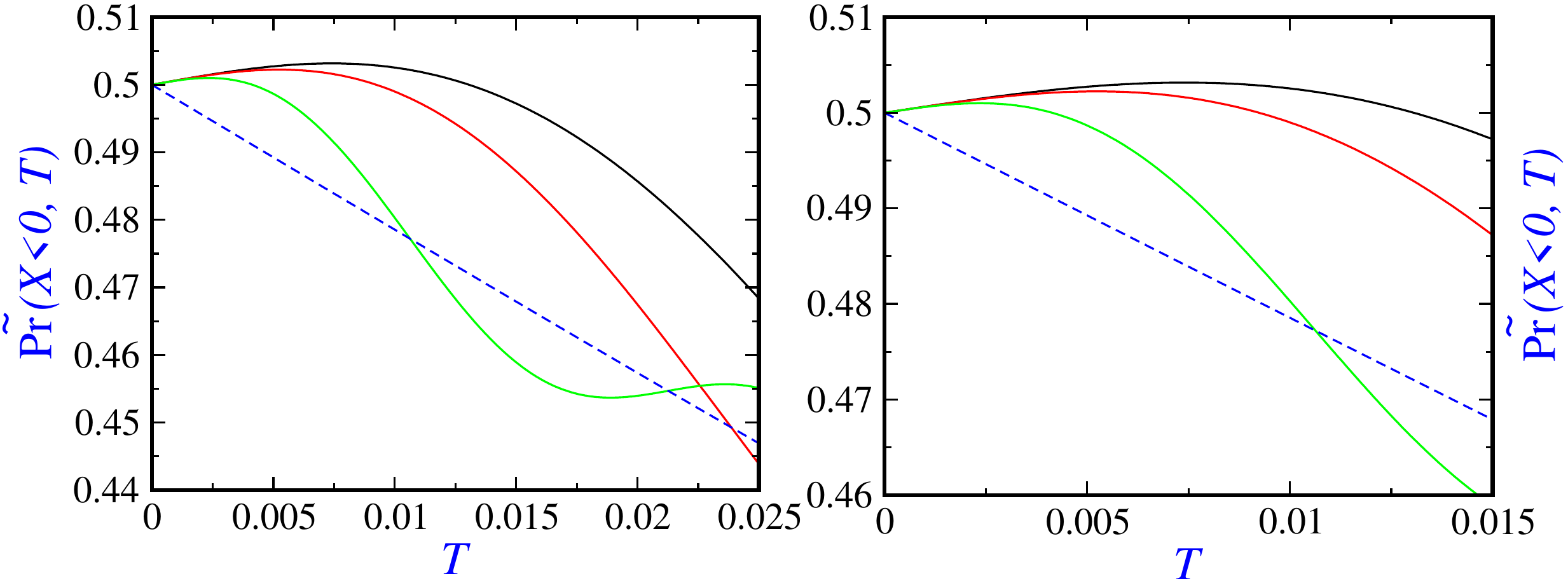}
		\caption{
			Probability of remaining in the negative half-space $ x< 0 $ given by Eq. (\ref{eq: prob_x<0}) for different dynamical regimes: quantum 
			(black), non-classical (red and green with $\ep=0.5$ and $\ep=0.1$, respectively) and classical (blue). The right panel is a close-up of 
			the left one around the origin.
		}
		\label{fig: Probt} 
\end{figure}

In figure \ref{fig: Probt}, we have plotted the probability in the negative half-space in terms of time for different regimes. 
For numerical computations, we have chosen $ P_{0a} = 14 $, $ P_{0b} = 3 $, $\alpha = 1.9 $ and $ \theta = \pi $. 
For these values, $\tilde{Pr}(p<0) $ is of the order of $ 10^{-10} $ for the quantum regime and even lesser for the remaining regimes.
The probability of backflow is thus observed for all non-classical regimes $ \ep \neq 0 $ except the classical one.

	%
	%

\section{Intrinsic decoherence in dissipative backflow}

In the Caldirola-Kanai framework for dissipative dynamics coming from an effective Hamiltonian, the time-dependent Schr\"{o}dinger 
equation reads as
\begin{eqnarray} \label{eq: CK}
	i \hb \frac{\pa }{\pa t} \psi(x, t)  &=& \bigg[ - e^{-2\ga t} \frac{\hb^2}{2m} \frac{\pa^2}{\pa x^2}  + e^{2\ga t} V(x) \bigg] \psi(x, t)  ,
\end{eqnarray}
where $\ga$ is the dissipation rate or friction coefficient.
Then, the scaled transition equation in this context is obtained as \cite{MoMi-AP-2018, MoMi-JPC-2018}
\begin{eqnarray} \label{eq: scaled_CK}
	i \thb \frac{\pa }{\pa t} \tpsi(x, t)  &=& \bigg[ - e^{-2\ga t} \frac{\thb^2}{2m} \frac{\pa^2}{\pa x^2}  + e^{2\ga t} V(x) \bigg] \tpsi(x, t)  .
\end{eqnarray}
In this framework, the continuity equation again has the form (\ref{eq: con_eq}) but with 
%
%
%
\begin{eqnarray} \label{eq: pcd_CK}
	\tj(x, t) &=& \frac{\thb}{m} \im \left\{ \tpsi^* \frac{\pa \tpsi}{\pa x}  \right\} e^{-2\ga t}   ,
\end{eqnarray}
as the probability current density.
The propagator of  the free particle in the context of dissipative transition equation is given by \cite{MoMi-JPC-2018}
\begin{eqnarray} \label{eq: free_CK_propagator}
		\tg_{\fr}(x, x'; t) &=& \sqrt{ \frac{m}{2\pi i \thb \tau(t) } }
		\exp \left[ \frac{i m}{2 \thb \tau(t)} (x-x')^2 \right]  ,
\end{eqnarray}
where
\begin{eqnarray}
		\tau(t) &=& \frac{ 1 - e^{- 2 \ga t} }{ 2 \ga }  .
\end{eqnarray}
	%
By introducing dimensionless time and relaxation coefficient according to
\begin{numcases}~
		T = \frac{t}{\nu}\\
		\Ga = \nu ~ \ga   ,
\end{numcases}
one obtains
\begin{eqnarray}
		\tau(t) &=& \nu \frac{ 1 - e^{- 2 \Ga T} }{ 2 \Ga } \equiv \nu ~ \uptau(T)  .
\end{eqnarray}
Now, by expressing the space coordinate versus $ \mu = \sqrt{\nu \hb /  m } $, one has
\begin{eqnarray} \label{eq: propagator2}
		\tg_{\fr}(x, x'; t) &=& \frac{1}{\mu} \frac{1}{ \ep^{1/4} } \frac{1}{\sqrt{ 2\pi i ~ \uptau(T) } }
		\exp \left[ \frac{i}{2 \sqrt{\ep} \uptau(T)} (X-X')^2 \right] \equiv \frac{1}{\mu} \tG_{\fr}(X, X'; T) \nonumber \\
\end{eqnarray}
for the propagator and for the time-evolved wave function
\begin{eqnarray}
		\tPsi(X, T) &=& \int_{-\infty}^{\infty} dX' ~ \tG_{free}(X, X'; T) ~ \tPsi(X',0)   .
\end{eqnarray}
Comparison of Eq. (\ref{eq: propagator2}) with the corresponding one for the non-dissipative case shows that previous results are 
valid replacing $ T $ by $ \uptau(T) $.

\begin{figure}  
		\centering
		\includegraphics[width=12cm,angle=-0]{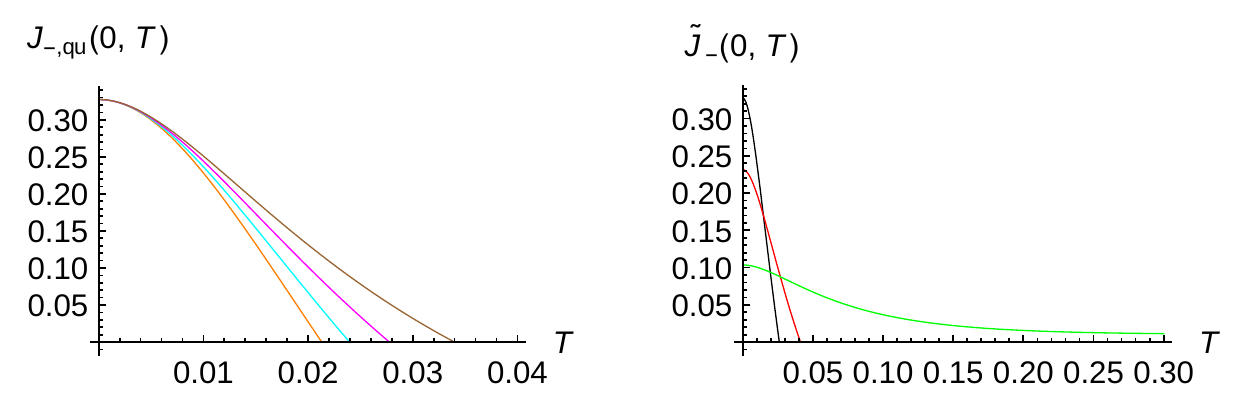}
		\caption{
			Negative part of the probability current density at the origin for different values of $\Ga$ in the quantum regime (left panel). 
			In the right panel, the same function is plotted for different non-classical regimes at $ \Ga = 8 $. The initial wave function is 
			given by Eq. (\ref{eq: Phi_BM}). Colour codes in the 
			left panel are: $ \Ga = 0 $ (orange), $ \Ga = 5 $ (cyan), $ \Ga = 10 $ (magenta) and $ \Ga = 15 $ (brown); and in the
			right panel:  $ \ep = 1 $ (black), $ \ep = 0.5 $ (red) and $ \ep = 0.1 $ (green).
		}
		\label{fig: plotPCDgama}
\end{figure}

For an illustration, in figure \ref{fig: plotPCDgama}, the negative part of the probability current density Eq. (\ref{eq: jneg})
at the origin versus time for the quantum regime with different values of $\ga$ (left panel) and for different non-classical regimes with 
a given value of $\ga$ (right panel) are shown. The initial wave function is given by Eq. (\ref{eq: Phi_BM}).  
As expected, the amount of backflow increases with damping in the quantum regime. 
On the contrary, in the quantum to classical transition, 
the negativity of the probability current density is again reduced.

	%
	%

\section{Conclusions}	

Along this work we have emphasized the importance of the intrinsic decoherence in backflow. This effect is gradually suppressed as the 
the corresponding process is developing. The emergence of classical mechanics or classical limit is quite similar irrespective of the 
theoretical method used, namely: the generalized von Neumann equation proposed by Milburn and the linear scaled wave equation 
which itself is equivalent to the non-linear transition wave equation.  By considering various examples, we have shown that the
intrinsic decoherence diminishes both the backflow amount and also its duration. As far as we know, 
this is the first time backflow has been considered in the most general formulation of quantum mechanics in terms of density matrices instead 
of wavefunctions and for isolated systems. Our results show that in the quantum-to-classical transition, this effect attenuates and completely 
disappears in the classical regime. This confirms us that backflow is a non-classical effect occurring in all non-classical dynamical  
regimes, including quantum mechanics. This should not be confused with the fact that this effect is seen even for  classical 
waves \cite{BiBiAu-JPA-2022}. The important point is that the backflow is an interference phenomenon which itself has a wave 
nature \cite{Gu-PRA-2019, Gu-PRR-2020}. Thus, only quantum particles with their dual character can exhibit this wave phenomenon. 
This situation is quite similar to tunnelling: while the effect is considered genuinely quantum mechanical, its classical optical analogue, 
frustrated total reflection, exists \cite{Krane-book-2012}. 
Another important point to be stressed is that the linearity of the wave equation is not necessary to see this wave phenomenon. Actually,  
as also shown here,  a nonlinear Schr\"odinger wave equation reproduces all the features of linear quantum  mechanics \cite{RiScVaBa-PRA-2014}.
Moreover, one of the main messages we want to convey is that the route to the classical limit is not unique i.e., there is not only one way to classicality. 
Intrinsic decoherence can also have several sources depending on the starting point. Even more, for completeness, a parallel theoretical analysis has been carried out in terms of the Linblad equation. Finally, it is also shown that the appearance of classicality through dissipative dynamics can not be comparable to the previous routes used.

Backflow has already been studied for both distinguishable \cite{MoMi-EPJP-2020} and  indistinguishable particles \cite{MoMi-RP-2020} 
in the context of the CK approach. In this work, we have carried out the same analysis but within the scaled CK equation 
\cite{MoMi-AP-2018, MoMi-JPC-2018}.  Previously, a study was carried out within the Caldeira-Leggett formalism \cite{MoMi-EPJP-2020} 
by including also temperature but then it was argued that the corresponding results did not imply quantum backflow 
\cite{MoMi-EPJP-2020-Era} due to the fact that the positive momentum distribution  along time was not maintained. 
With intrinsic decoherence, extension to important aspects such as, for example, its effect on  backflow for two-identical-particle systems 
and the interference pattern for a superposition of two wave packets are interesting. Work in this direction is now in progress.

\section{Acknowledgments}

SVM acknowledges support from the University of Qom. 
SMA would like to thank support from Fundaci\'on Humanismo y Ciencia.
Furthermore, we thank the anonymous reviewers for their careful reading of our manuscript and their many insightful comments and suggestions.


\end{document}